\documentclass[a4paper,11pt]{article}
\pdfoutput=1 

\usepackage{jcappub} 

\usepackage[T1]{fontenc} 

\usepackage{subcaption}
\usepackage{comment}
\usepackage{multirow}

\usepackage{xcolor}

\title{\boldmath Dynamical Friction due to fuzzy dark matter on satellites described by axisymmetric logarithmic potentials }

 \author{Andreas Vitsos}
 \author{and Konstantinos N. Gourgouliatos}
\affiliation{University of Patras, Department of Physics,\\26504, Rio, Patras, Greece}

\emailAdd{up1055693@upnet.gr}
\emailAdd{kngourg@upatras.gr}

\abstract{
A plausible dark matter candidate is an ultralight bosonic particle  referred to as fuzzy dark matter. The equivalent mass-energy of the fuzzy dark matter boson is $\sim 10^{-22}$eV and has a corresponding de Broglie wavelength of kiloparsec scale, thus exhibiting wave behaviour in scales comparable to a galactic core, which could not appear in conventional cold dark matter models. The presence of fuzzy dark matter in galactic clusters will impact the motion of their members through dynamical friction. In this work, we present simulations of the dynamical friction on satellites traversing an initially uniform fuzzy dark matter halo. We focus on the satellites whose shapes are beyond spherical symmetry described by ellipsoidal and logarithmic potentials. We find that the wakes created on the fuzzy dark matter halo due to the passage of such satellites are qualitatively different from those generated by spherically symmetric ones. Furthermore, we quantify the dynamical friction coefficient for such systems, finding that the same satellite may experience a drag differing by a factor of $5$ depending on its ellipticity and the direction of motion. Finally, we find that the dynamical friction time-scale is close to Hubble time, assuming a satellite of $10^{11}$M$_{\odot}$ traversing at $10^{3}$km/s a FDM halo whose mean density is $\sim 10^6$M$_{\odot}$kpc$^{-3}$.}

\begin{document}
\maketitle
\flushbottom

\section{Introduction}
\label{sec:intro}

According to the standard Cosmological model of Cosmology \cite{2020A&A...641A...6P}, dark matter (DM) constitutes approximately 27$\%$ of the content of Universe, yet, it has not been possible to decipher its nature. Since early observations of galactic velocities \cite{1933AcHPh...6..110Z}, the "missing mass" of galaxies \cite{1940ApJ....91..273O}, and their rotation curves \cite{1970ApJ...159..379R,1973A&A....26..483R} the current $\Lambda$ Cold Dark-Matter model, ($\Lambda$CDM) has been in agreement with the consensus of Cosmological observations \citep{2015PNAS..11212243B}, including the Cosmic Microwave Background Radiation and the large scale structure of the Universe \cite{2020A&A...633L..10T} with mounting evidence from independent sources supporting the presence of DM, from primordial abundances \cite{2012MNRAS.425.2477P} to gravitational lensing \cite{2009ApJ...703.2232S}. Despite these successes, the search for a particle candidate, such as the WIMP particle, has been fruitless so far either through the experimental particle physics route \cite{2015PNAS..11212256P} or observational astrophysics \cite{2015PNAS..11212264F}. Several candidates have been proposed for the nature of DM in the form of particles, i.e.~axions \cite{1983PhLB..120..127P}, neutrinos \cite{1986PhLB..167..295F,2021PhRvL.127w1801D}); macroscopic objects i.e.~primordial black holes \cite{2010JCAP...04..023F}, massive compact halo objects \cite{1991ApJ...366..412G}; or even modifications in the theory of gravity and dynamics \cite{1983ApJ...270..365M}. 

A generic feature of cold DM models is that they predict a cuspy core for DM haloes. Cuspy cores correspond to a profile where the density diverges in the centre of the DM halo \cite{1996ApJ...462..563N}. This can be appreciated intuitively as cold DM has no pressure to counteract collapse. Initially motivated by this issue, Fuzzy Dark Matter (FDM) tackles successfully this problem by postulating ultra-light scalar particles as DM candidates \cite{Baldeschi,History,Hu}. Apart from resolving the cuspy-core problem, it also addresses other discrepancies between observation and theory regarding dwarf galaxies in the context of the missing satellite problem \cite{Hu}. While the hypothesis of an ultra-light particle seems to face some challenges \cite{Burkert,2020ApJ...893...21S} the FDM model is a noteworthy candidate for DM. FDM belongs to the lower end of the ultralight DM regime. It consists of cold scalar particles with a mass $m \sim 10^{-22}$eV, for comparison the lightest DM candidate, the axion, is searched through experiments at a range of $10\mu $eV$ - 1 $meV \cite{Axionmass}. The FDM particle is postulated to be a boson forming Bose-Einstein condensates. An important property of FDM is that the de Broglie wavelength is of the order $\lambda_{dB} \sim 1kpc$. As a result there are quantum effects exhibited on astrophysical scales, namely quantum pressure that counteracts the gravitational collapse of the DM halo, thus resolving the cuspy-core problem.

A FDM condensate will cause a distinctive behaviour to massive objects whose orbits lie within its halo. In general, a massive object passing through a cloud of smaller bodies experiences a drag due to the mutual gravitational interaction. This effect is called dynamical friction \cite{1943ApJ....97..255C,1943ApJ....97..263C,1943ApJ....98...54C}. The interaction between the satellite and the FDM condensate also results in the creation of wakes on the DM halo. Moreover, the fluid nature of FDM leads to the formation of smaller-scale structures such as vortices \cite{hui2021wave} depending on the properties of the details of the FDM model employed. Such phenomena related to a FDM halo have been studied analytically and numerically, recently, considering the drag on a point mass and a spherically symmetric extended mass described by the Plummer potential \cite{Lancaster,traykova2021dynamical,Hartman}. As satellites are not, in general, spherically symmetric, an unresolved yet question, is the  dynamical friction exerted by an FDM halo on objects described by non spherically-symmetric potentials and the formation of  wakes to the FDM halo. Therefore, in this work we study the interaction of objects who are described through non-axisymmetric logarithmic potentials with a FDM halo. To this end we simulate numerically the evolution of such system for a broad range of parameters utilizing the formulation described in \cite{Lancaster}, and we provide a quantitative estimate on the impact of FDM to systems for which the velocities are known.

The structure of the paper is as follows. In section \ref{MATH_FORM} we present the governing equation for FDM and the numerical scheme employed for the study of the problem. In section \ref{SIMULATIONS} we present the numerical simulations performed. Section \ref{RESULTS} contains the results of the simulations. We discuss our results in section \ref{DISCUSSION}. We conclude in section \ref{CONCLUSIONS}

\section{Mathematical Formulation}
\label{MATH_FORM}

\subsection{Schr\"odinger-Poisson Equation}

The quantum mechanical behaviour of FDM which, due to the very low particle mass extends to macroscopic scales, dictates the use of the Schr\"odinger equation for its description. As the main driving force in this system is gravity, we need to include in our calculation the gravitational potential, which is determined by the Poisson equation. Combining these two equations we obtain the Schr\"odinger-Poisson equation, which is the governing equation of the system \cite{2018PhRvD..97h3519M}.  

Consider the Schr\"odinger's equation for a FDM condensate: 
\begin{equation}
    i \hbar\frac{\partial\psi}{\partial t} = \left[ -\frac{\hbar^2}{2m}\nabla^2 + m(U_s+U)\right] \psi\,,
        \label{eq:Schrodinger}
\end{equation} 
where $\psi$ is the condensate's wave-function, $\hbar$ is the reduced Planck’s constant, $m$ is the mass of the FDM particle, $U_s$ is the satellite potential and $U$ is the  potential term confining the condensate. The potential is due to the gravitational field, thus we use the Poisson equation to obtain a relation between the mass density $\rho_U$ associated to the potential $U$:
\begin{equation}
\nabla^2 U = 4 \pi G \rho_U\,,
\label{eq:Poisson}
\end{equation}
where $G$ is Newton's gravitational constant.

The combination of Sch\"odinger and Poisson equation into a system of coupled partial differential equations is achieved through the Madelung formalism \cite{1927ZPhy...40..322M}. The wave-function can be written in terms of an amplitude and a complex phase: 
\begin{equation}
    \psi = \sqrt{\rho} e^{i\theta}\,,
    \label{eq:wavefunction}
\end{equation}
where the $\rho$ is the density of the FDM. Moreover, we can relate the FDM velocity $\mathbf{u}$ to the phase of the wave function:
\begin{equation}
  \mathbf{u}=\frac{\hbar}{m}\nabla \theta\,.  
\end{equation}
Thus, the combination of equations \ref{eq:Schrodinger}, \ref{eq:Poisson} and \ref{eq:wavefunction}, along with an appropriate set of initial conditions allows for the solution of a self-gravitating FDM system \cite{Mocz,2019PhRvL.123n1301M,2020MNRAS.494.2027M}.

\subsection{Dynamical Friction}

Next we include the role of the collective drag of FDM halo onto a body travelling through this halo, in a straight-line orbit. A body moving through a FDM condensate will experience effects of dynamical friction, that also alter the density of the medium. We will refer to the body moving within the FDM medium as satellite. While the satellite traverses through the medium a wake and more complex, physically notable structures, may appear around it. 

While a self-consistent solution for the entire satellite-FDM halo system is technically feasible, we are going to approximate the problem of a moving satellite through an FDM halo as follows. We select a frame of reference that is co-moving with the satellite. The satellite corresponds to a given potential function $U_s$ which is set to a particular profile and does not evolve with time, as we consider the impact of the FDM  onto the satellite mass distribution to be minimal compared to the effect the satellite has onto the FDM halo. We consider a FDM halo, whose initial density and velocity are constant. Subject to these requirements we integrate in time equation \ref{eq:Schrodinger}, while as the gravitational potential depends only on the mass profile of the satellite, there is no need to integrate equation \ref{eq:Poisson}, as it is satisfied self-consistently for a satellite with an appropriate density profile, which is significantly higher than that of the FDM halo. This approximation is valid for regions where the satellite is significantly denser than the halo. A satellite traversing the outskirts of a halo fulfills this requirement, however once it approaches the core, one needs to account for the total density distribution in the integration of the Poisson and Schr\"odinger equation. The approximation of a straight-line path holds provided the orbit radius is sufficently long. 

To quantify the impact of the dynamical friction we are going to utilise a dimensionless dynamical coefficient given by the ratio of a reference force and the dynamical friction force \cite{Lancaster}. The reference force is given in terms of the characteristic quantities appearing in the system, defined as follows:
\begin{equation}
    F_{ref} \equiv 4 \pi \bar{\rho} \left(\frac{GM}{u_{rel}}\right)^2\,,
\end{equation}
where $M$ is the mass of the satellite, $u_{rel}$ is the satellite velocity relative to the background and $\bar{\rho}$ is the average density of the FDM. We further define the dimensionless dynamical friction coefficient as follows:
\begin{equation}
    C_{rel} \equiv \frac{F_{DF}}{F_{ref}}\,,
\end{equation}
where $F_{DF}$ is the dynamical friction force which is evaluated as follows:
\begin{equation}
F_{DF} = -\bar{\rho} \int_{V} dV \frac{\rho -\bar{\rho}}{\bar{\rho}} \frac{\bf u_{rel}}{|\bf{u_{rel}}|} \cdot \nabla U\, 
\label{eq:FDF}
\end{equation}
Note that the sign is negative, as we refer to the force exerted by the halo to the satellite.
Thus, after solving Schr\"odinger's equation and finding the wavefunction $\psi$, and consequently the FDM density, we can directly evaluate the integral above and determine the force arising from the dynamical friction.

\subsection{Numerical Integration}

We integrate numerically equation \ref{eq:Schrodinger} applying the "Kick-Drift-Kick" technique, a leapfrog-like integrator, using appropriate Fourier and inverse Fourier transformations based on the framework developed in \cite{Mocz}. Schr\"odinger equation has a kinetic and a potential part which are associated to the Laplacian and the potential operator, respectively, acting onto the wave-function. The potential term is evaluated by two half kick-steps of the wave-function in real space, intermediated by a full step of the kinetic term in Fourier space, as follows:
\begin{subequations}
 \begin{align} \label{eq:Kick-Drift-Kick}
 &\quad \phantom{={}} \psi \leftarrow \exp \left[-i\frac{\Delta t}{2} \frac{m}{\hbar} U_s\right] \psi,\\
 &\quad \phantom{={}} \psi \leftarrow \text{ifft} \left\{\exp\left(-i k^2 \Delta t \frac{\hbar}{2m}\right) \text{fft}\left[\psi\right]\right\},\\
 &\quad \phantom{={}} \psi \leftarrow \exp \left[-i\frac{\Delta t}{2} \frac{m}{\hbar} U_s\right] \psi,
 \end{align}
 \end{subequations}
where fft and ifft denote the fast Fourier and the inverse fast Fourier transformation, respectively, $\Delta t$ is the time step, and $k$ is the wavenumber. This process is repeated, providing the time evolution of the wavefunction describing the FDM condensate.

For the integration, we have developed a code in Python, using the built-in routines of the package NumPy for the implementation of the fast fourier transformation and its inverse. We have adopted a Cartesian grid, where the relative motion between the satellite and the FDM condensate is along the $z$ axis. We have studied 2-D and 3-D systems with a typical mode resolution in the Fourier space of $N=256$ in 3-D, and $N=512$ in the 2-D runs. The time-step was set to $\Delta t = 0.005$, ensuring numerical convergence in all runs. We have further run a selected number of cases in higher resolution to ensure the numerical convergence of the system, which was the case, indeed. The typical size of the integration box for both the 2-D and the 3-D runs is $L=50 \pi$, unless specified otherwise. Regarding the boundary conditions, we have applied periodic ones on every boundary of the domain. In all runs we initialise our domain with a FDM background of uniform density, thus $\|\psi\|=1$, and a uniform velocity along the $z$ direction. 

Regarding the evaluation of the dynamical friction coefficient, we perform the following process. The potential, mass, and velocity of the satellite are known and do not evolve with time for a given simulation. The numerical solution of the Schr\"odingr equation \ref{eq:Schrodinger} provides the form of $\psi$. Therefore, we integrate numerically equation \ref{eq:FDF} and evaluate the dynamical friction acting onto the satellite.

  \section{Simulations}
 \label{SIMULATIONS}

Several previous works have focused on the effect of FDM dynamical friction onto a point mass, i.e.~black holes, for which analytical solutions also exist \cite{2022PhRvD.105f3523W,2022PhRvD.105h3008V}, and extended distributions of masses motivated by systems such as the Magellanic Clouds, Fornax globular clusters and the Sagittarius streams \cite{Lancaster,Hartman}. In these studies, the extended masses are described through the spherically symmetric, Plummer potential \cite{1911MNRAS..71..460P} given by the following expression
\begin{equation}
    U_s = -\frac{GM}{\sqrt{r^2+R_c^2}}\,,
    \label{eq:Plummer}
\end{equation}
where $M$ is the total mass, $R_c$ the radius of the core and $r$ is the radial coordinate measured from the centre of the source, and for large $r$ asymptotically, it reduces to a point mass potential. 

Yet, numerous systems that are influenced by dynamical friction, residing in extended FDM concentrations, may deviate from spherical symmetry and be described by more generic potentials. Because of this, the main focus of the present work is to study axisymmetric logarithmic potentials \cite{1980ApJ...238..103R,1981MNRAS.196..455B}, with the inclusion of a core radius, which have been used extensively for such systems. The basic potential expression for these sources is given by the following mathematical relation:
\begin{equation}
    U_s = \frac{v_c^2}{2}\ln\left(R^2+\frac{y^2}{b_y^2}+R_c^2\right)
\end{equation}
where $R^2=x^2+z^2$, thus the system is axially symmetric round the $y$-axis, $b_y$ is the ellipticity parameter and $v_c$ is a velocity equal to the orbital circular velocity for $r\gg R_c$. The satellite moves along the $z$ direction with respect to the FDM condensate, thus the satellite axis of symmetry is normal to the direction of motion. We have also studied another family of potentials where the axis of symmetry is parallel to the direction of motion. These systems are described by the following potential:
\begin{equation}
    U_s = \frac{v_c^2}{2}\ln\left(R^2+\frac{z^2}{b_z^2}+R_c^2\right)\,,
\end{equation}
where the axial radius now is given by $R^2=x^2+y^2$. We note that the physically meaningful values of ellipticity parameter lie within the range  
\begin{equation}
    \frac{1}{\sqrt{2}}<b_{y,z}<1.08\,,
\end{equation}
as values outside this range yield negative mass densities \cite{GalacticDynamics,1993MNRAS.260..191E}.  

We have further explored, the two-dimensional analogue of the above system
\begin{equation}
    U_s = \frac{v_c^2}{2}\ln\left(\frac{y^2}{b_y^2}+\frac{z^2}{b_z^2}+R_c^2\right)\,,
    \label{eq:log2D}
\end{equation}
where we set either $b_y,~b_z$ to unity and vary the other within the range of permitted values. These models are lower-dimensionality simplifications of the full three-dimensional ones and permit substantially faster numerical simulations, allowing us to obtain physical insight for the more complex three-dimensional cases. The full range of parameters we have explored is summarised in table \ref{Table1}.
\begin{table}[t]
 \centering
 \begin{tabular}{|c|c|c|c|c|c|c|}
 \hline
Potential Type & Figure & $M_Q$ &  $by$ &$b_z$ & $v_c$ & $R_c$ \\

\hline
& & $1$ &   & & & \\ 
& & $1/2$ &     & & & \\ 
Spherical (Plummer)&  Figure 1 & $1/4$ & - & - & - & 0.5 \\ 
& & $1/8$ &    & & & \\ 
& & $1/16$ &   & & & \\ 
\hline

\multirow{10}{*}{Logarithmic 3-D} & \multirow{5}{*}{Figure 2} & \multirow{10}{*}{1/10}  & \multirow{5}{*}{-} & 0.707 & \multirow{10}{*}{1.0} & \multirow{10}{*}{0.9} \\ 

& & & & 0.8 & & \\

& & & & 0.9 & & \\

& & & & 1.0 & & \\

& & & & 1.08 & & \\

\cline{2-2}
\cline{4-5}

& \multirow{5}{*}{Figure 3} & & 0.707 & \multirow{5}{*}{-}  & & \\

& & & 0.8 & & & \\

& & & 0.9 & & & \\

& & & 1.0 & & & \\

& & & 1.08 & & & \\
\hline

\multirow{8}{*}{Logarithmic 2-D} & \multirow{4}{*}{Figure 4} & 1/2 & \multirow{4}{*}{0.9} & \multirow{4}{*}{1} & \multirow{8}{*}{1.0} & \multirow{8}{*}{0.9}\\ 

& & 1/4 & & & & \\

& & 1/8 & & & & \\

& & 1/16 & & & & \\

\cline{2-5}

& \multirow{4}{*}{Figure 5} & 1/2 & \multirow{4}{*}{1} & \multirow{4}{*}{0.9} & & \\ 

& & 1/4 & & & & \\

& & 1/8 & & & & \\

& & 1/16 & & & & \\
\hline

 \end{tabular}
 \caption{Summary of simulations performed.}
 \label{Table1}
 \end{table}

The main physical parameters included in this study are the mass of the FDM particle $m$, the mass of the object orbiting within the FDM condensate $M$ and its velocity. Combining these quantities, along with the Planck's constant $\hbar$ and $G$ can be combined into a dimensionless parameter, the quantum Mach number \cite{Hui}:
\begin{equation}
     M_Q \equiv \frac{u_{rel}\hbar}{GMm}\,,
\end{equation}
which, for the typical values encountered here, scales as follows
 \begin{equation}
     M_Q = 0.45 \left(\frac{u_{rel}}{100 {\rm km \ s^{-1}}}\right) \ \left(\frac{m}{10^{-22}{\rm eV}}\right)^{-1} \ \left(\frac{M}{10^9 M_{\odot}}\right)^{-1}\,.
     \label{eq:Mq}
 \end{equation}
As expected, faster moving systems have higher quantum Mach numbers, while more massive ones have lower $M_Q$. Motivated by the typical values for a sample of galaxies, using measurements from observations \cite{Simbad,SB1,SB2,SB3,SB4,SB5,SB6,SB7,SB8,SB9,SB10}, we focus on values of $M_Q<1$, the so-called ``classical" regime, which can be simulated with the kick-drift-kick method. We note that the logarithmic potential is parametrised through the velocity $v_c$ which scales with the mass of the satellite as $v_c^2 \sim \frac{GM}{R_C}$, thus we can use the expression $GM \sim R_C v_c^2$. The typical core radius we have simulated  lies within the range $R_C \in [0.5,\,1]$ which corresponds to a physical size of $\sim 1$kpc which is consistent the characteristics of the Milky-Way \cite{2016ARA&A..54..529B}. 
 
 \section{Results}
 
 \label{RESULTS}

\begin{figure}[tbp]
    \centering
    \includegraphics[width =\textwidth]{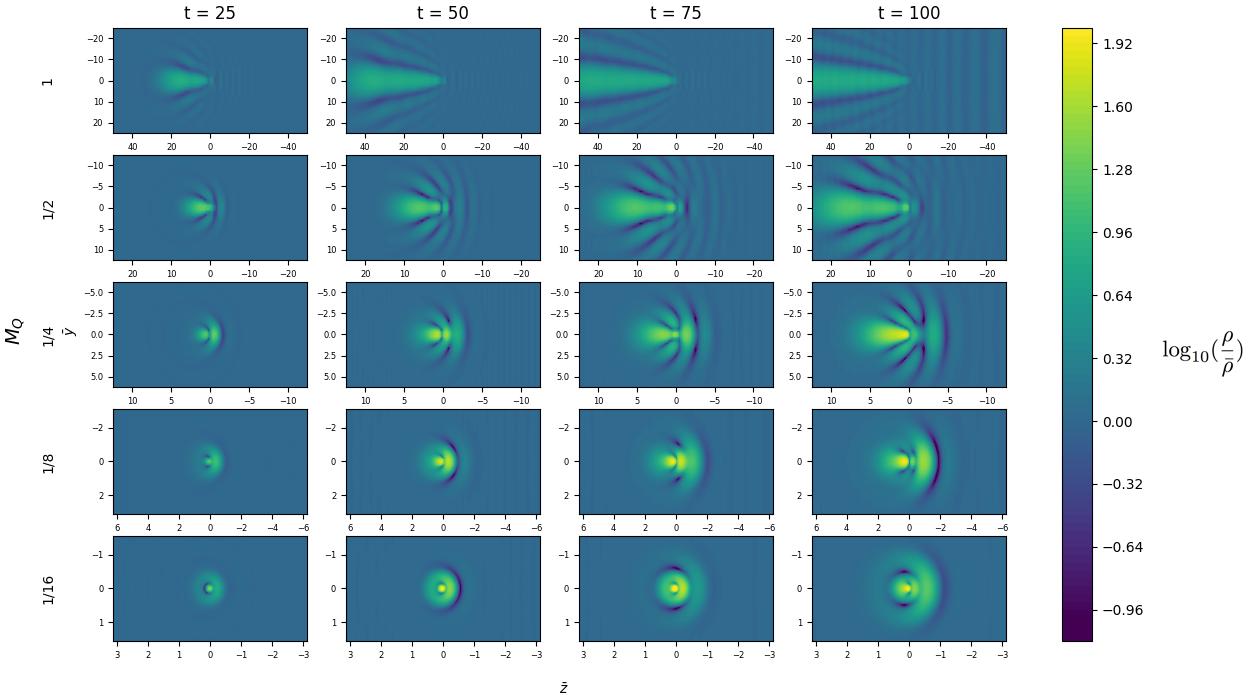}
    \caption{Snapshots of the FDM density contrast $a$ at various times for a spherical Plummer potential with $R_C=0.5$ travelling through a FDM for different values of $M_Q$, as indicated on each row. }
    \label{fig:2}
\end{figure}

We have performed a series of runs using the Plummer potential, equation \ref{eq:Plummer}, for various velocities parametrised with $M_Q$, and core size $R_c=0.5$ with successive snapshots shown in Figure~\ref{fig:2} zooming in the central region. A wake forms in the higher velocities simulated ($M_Q=1,\, 1/2,\, 1/4$, top three rows of Figure~\ref{fig:2}), similar to the point mass solution, cf.~Figure 2 of \cite{Lancaster}. The wake consists of an overdensity trailing the potential, which is responsible for dynamical friction. However, slower velocities ($M_Q=1/8,\, 1/16$) 
bottom two rows of Figure~\ref{fig:2}) lead to qualitatively different profiles, with smaller wakes trailing the satellite, and overdensities appearing ahead of it. In the lower velocity systems, the wake approximates a semi-circular shape, with both an overdensity and underdensity ahead of the satellite. This behaviour is in accordance with \cite{Lancaster,Hui} who found that the overdensity upstream of the satellite becomes suppressed for higher $M_Q$ numbers, due to FDM's wave nature. We further note that the wake of the perturbation in the subsonic regime has a characteristic wavelength that scales with the quantity $M_Q$, which is evident in Figure \ref{fig:2} rows 3-5.

\begin{figure}[tbp]
    \centering
    \includegraphics[width =\textwidth]{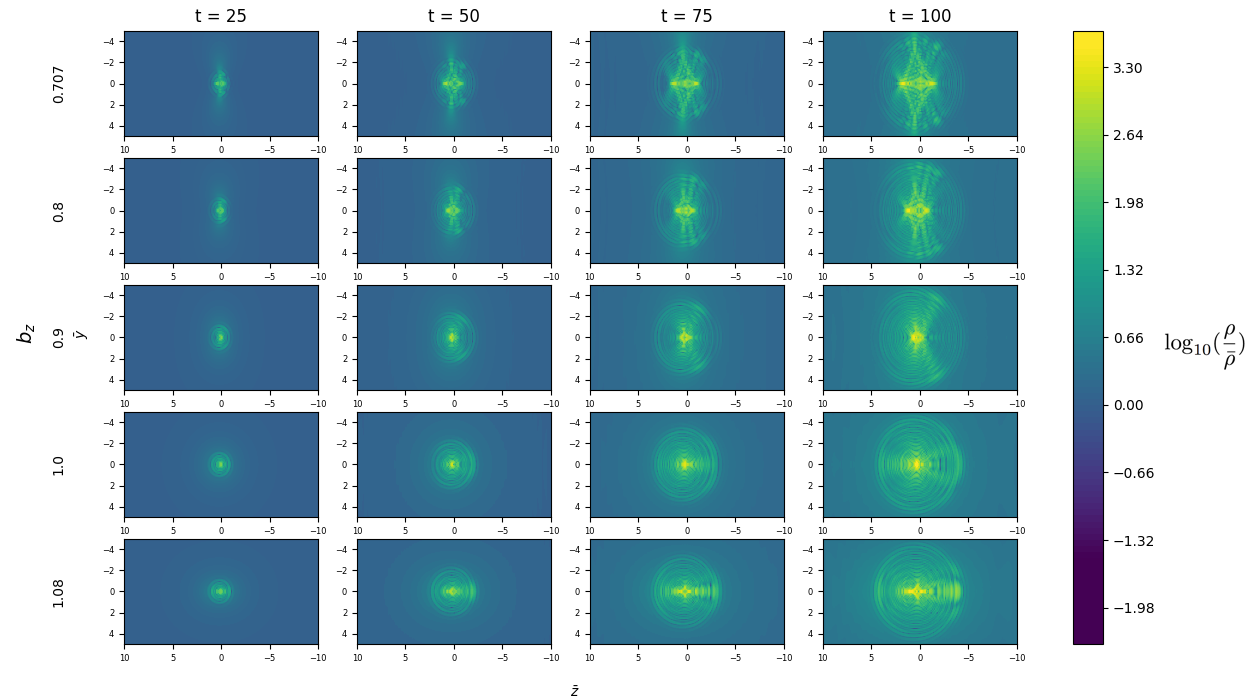}\caption{Snapshots of the FDM density ratio over the average density at various times for a logarithmic potential with $R_C=0.9$ and $M_Q=0.1$. The rows from top to the bottom have $b_z=0.707,\,0.8,\,0.9,\,1,\,1.08$, with the first three cases corresponding to oblate spheroid, the forth being spherically symmetric and the fifth prolate spheroid. The axis of symmetry of the ellipsoid is parallel to the direction of motion.}
    \label{fig:3}
 \end{figure}

Next, we have studied the axisymmetric logarithmic potential for a system whose symmetry axis is parallel to the direction of motion, Figure \ref{fig:3}, for various choices of the ellipticity parameter, while keeping the quantum Mach number $M_Q=0.1$ and $R_C=0.9$. The impact of the ellipticity parameter in the shape of the wakes created is evident. Oblate spheroids, i.e.~models with $b_z<1$, tend to create features which extend to larger distances on the $y$ axis, while a prolate spheroid ($b_z=1.08$) tends to create characteristic features extending in the $y$ direction, with an overdensity ahead of the of the direction motion. A spherically symmetric, logarithmic potential ($b_z=1$) fills in the continuum in terms of features between the oblate and the prolate configurations. 

\begin{figure}[tbp]
    \centering
    \includegraphics[width =\textwidth]{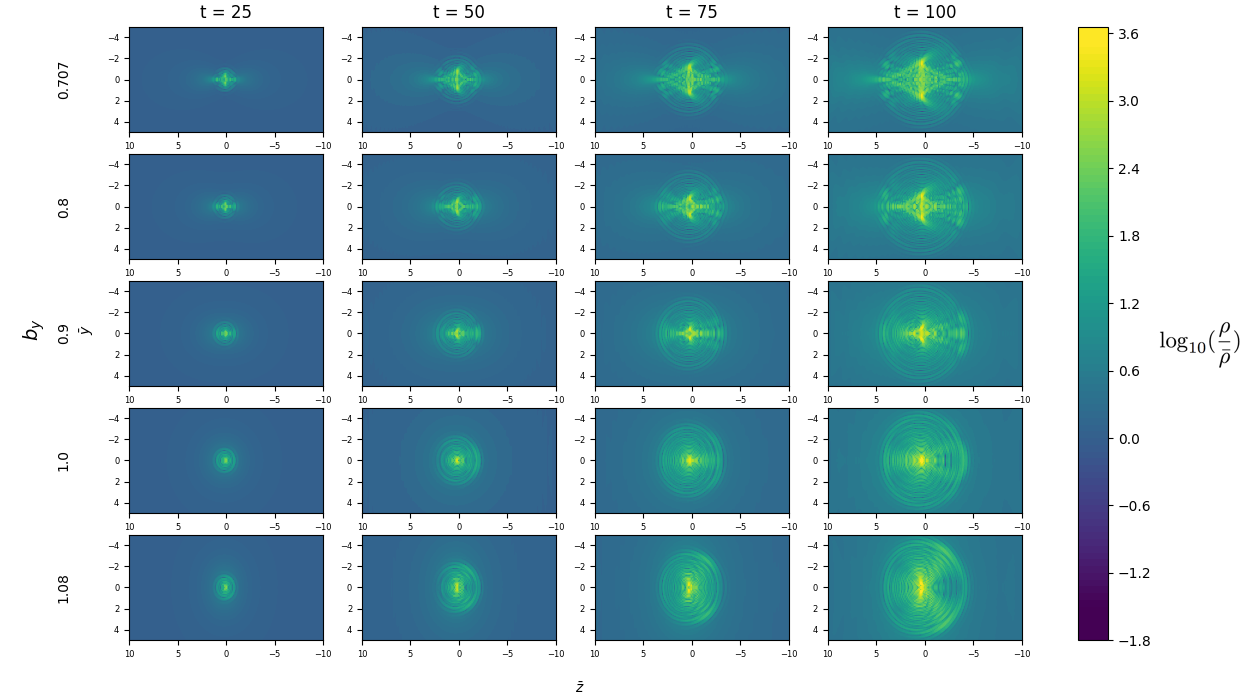}
    \caption{Snapshots of the FDM density ratio over the average density at various times for a logarithmic potential with $R_C=0.9$ and $M_Q=0.1$. The rows from top to the bottom have $b_y=0.707,\,0.8,\,0.9,\,1,\,1.08$, with the first three cases corresponding to prolate spheroid, the forth being spherically symmetric and the fifth oblate spheroid. The axis of symmetry of the ellipsoid is parallel to the direction of motion.}
    \label{fig:4}
 \end{figure}

Furthermore, we have simulated an axisymmetric logarithmic potential where the axis of symmetry is normal to the direction of motion, Figure \ref{fig:4}. We notice that the characteristic wake in this system is qualitatively different, especially for the oblate spheroid cases ($b_y=0.707,\, 0.8,\,0.9$). The spherically symmetric case $b_y=1$ is identical to the one with $b_z=1$, as expected, and the prolate spheroid ($b_y=1.08$) does not bear significant differences compared to the $b_z=1.08$ case.  We note that there is a significant difference between the  spherically symmetric cases of the logarithmic potentials and the Plummer one even when the velocities are similar ($M_Q=0.1$ and $M_Q=1/8$), see Figure \ref{fig:2} third row and Figures \ref{fig:3} and \ref{fig:4} fourth row. This difference is related to the fact that the logarithmic potential corresponds to a more extended mass distribution, as opposed to the Plummer one. Because of this, the latter leads to a more spherically symmetric configuration whereas the Plummer potential leads to more asymmetric one.

Next, we have studied the two-dimensional logarithmic models given by equation \ref{eq:log2D}, with $R_C=0.9$, $v_c=1$, exploring different values of $M_Q$ while keeping $b_z=0.9$ and $b_y=0.9$ respectively. The main feature of these simulations is the drastic dependence on the velocity, in particular the size of the affected region depends drastically on the velocity. In particular, for high velocities i.e.~$M_Q=0.5$ the affected region extends to a radius of $20$ units, while for lower ones, such as for $M_Q=0.0625$ the affected region has a radius of a few units of length. Overall, in  logarithmic potentials, the shape of the perturbation on the dark matter halo reflects the shape of the potential of the satellite. This is related to the fact that the velocities we are studying here are typically low, thus we do not expect to see wakes as in the models employing high values for $M_Q$ and the fact that the corresponding mass distribution is rather extended. 
 \begin{figure}[tbp]
    \centering
    \includegraphics[width =\textwidth]{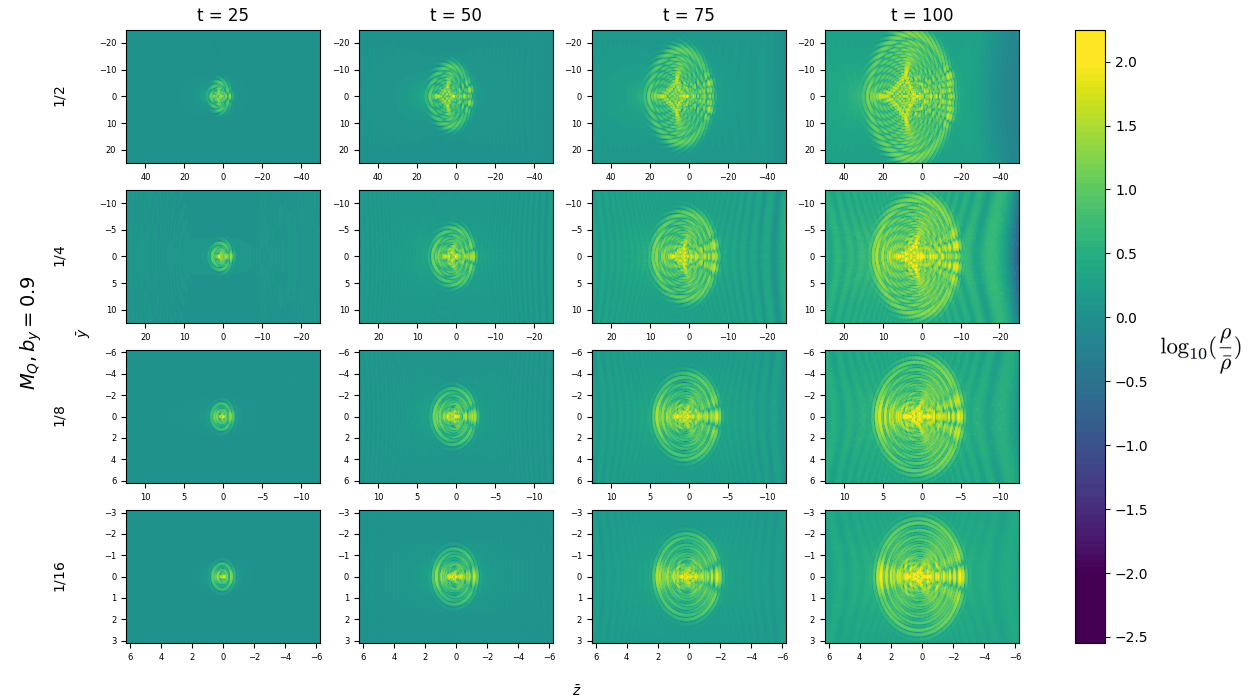}
    \caption{Snapshots of the FDM density over the average density at various times for a 2-D logarithmic potential with $R_C=0.9$ and $b_{y}=0.9$ for various values of $M_Q$.}
    \label{fig:5}
 \end{figure}
  \begin{figure}[tbp]
    \centering
    \includegraphics[width =\textwidth]{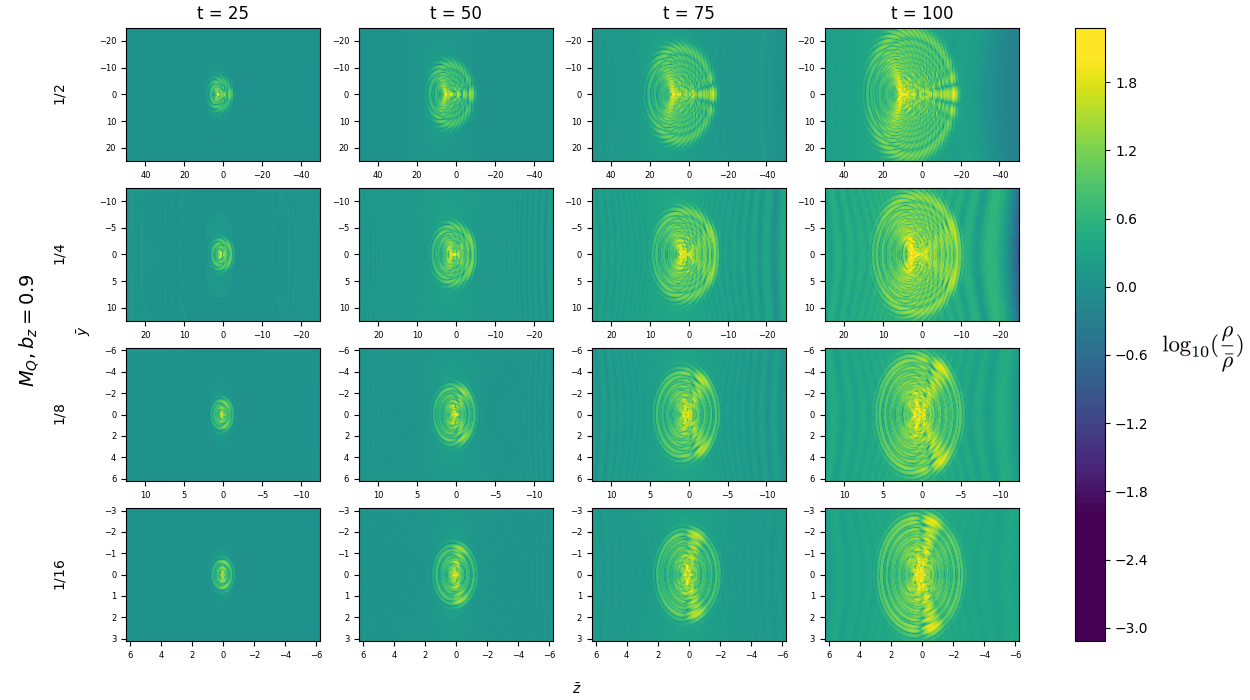}
    \caption{Snapshots of the FDM density over the average density at various times for a 2-D logarithmic potential with $R_C=0.9$ and $b_{z}=0.9$ for various values of $M_Q$.}
    \label{fig:6}
 \end{figure}

\begin{figure}[tbp]
    \centering
    \includegraphics[width =\textwidth]{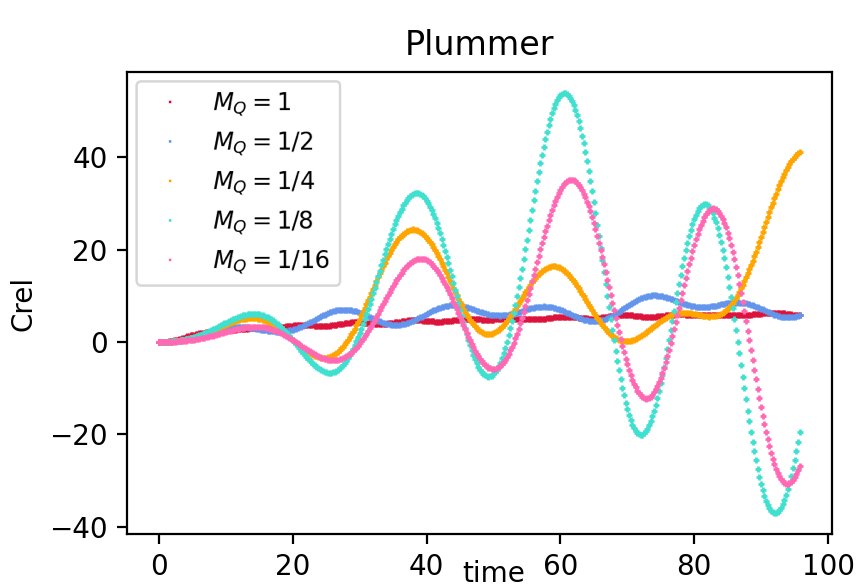}
    \caption{The dynamical friction coefficient $C_{rel}$ as a function of time for Plummer potentials for various values of $M_Q$}
    \label{fig:7}
 \end{figure}

We have evaluated the dynamical friction coefficient for the simulations corresponding to the potentials we have studied. In the system containing a satellite described by the Plummer potential, Figure \ref{fig:7}, we notice that the dynamical friction coefficient is lower for higher $M_Q$ values. Quite remarkably we notice that the coefficient becomes negative in the low $M_Q$ regime, and fluctuates drastically, between negative and positive values. This is attributed to the overdensities and underdensities that appear in the waves, which may in some cases accelerate the satellite instead of dragging it. We note however, that the oscillatory behaviour of $C_{rel}$ in the low $M_{Q}$ runs implies that the mean force will be somewhat lower than the maximum attained. Nevertheless, the overall trend is that for higher $M_{Q}$ the parameter $C_{rel}$ decreases. This is in accordance with the study of the $M_Q$ parameter space i.e.~Figure 7 of \cite{Lancaster} and Figure 6 of \cite{2021PhRvD.104j3014T},  where the dynamical friction force does not have a monotonic dependence with the velocity.

\begin{figure}[tbp]
    \centering
    a\includegraphics[width =0.45\textwidth]{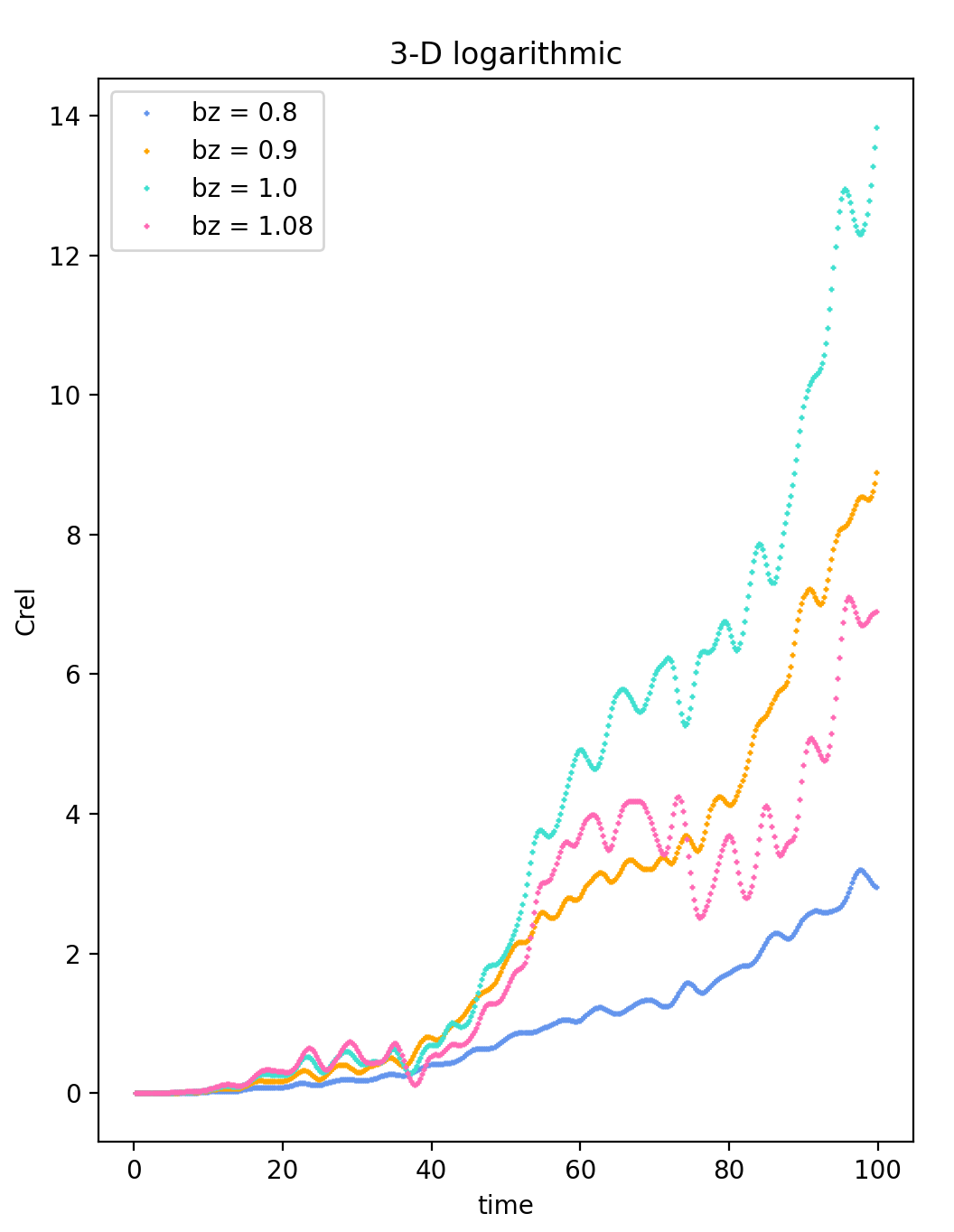}
        b\includegraphics[width =0.45\textwidth]{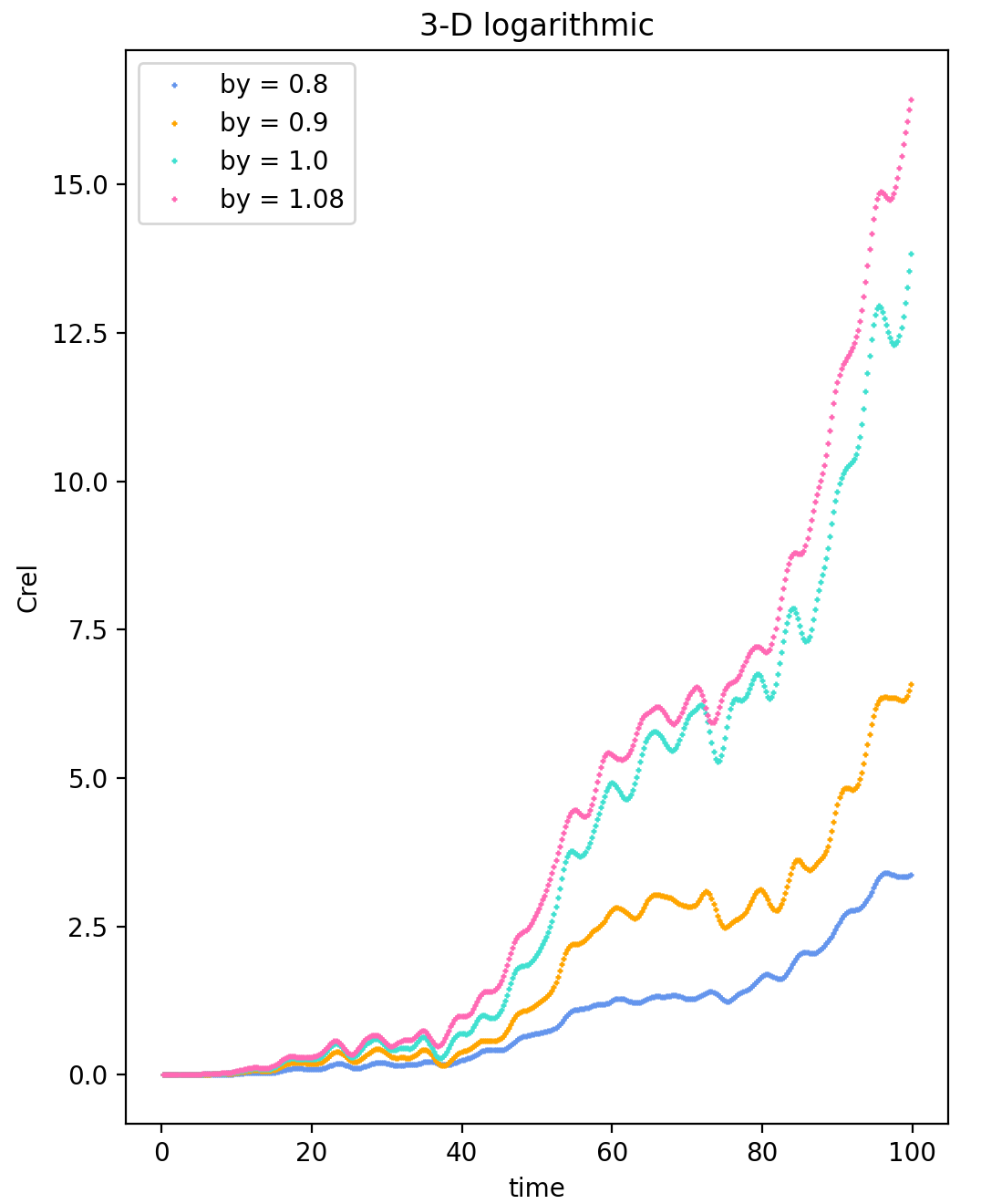}
    \caption{(a).The dynamical friction parameter $C_{rel}$ as a function of time for 3-D logarithmic potentials with various $b_z$, where the axis of symmetry is parallel to the direction of motion. (b) The dynamical friction coefficient $C_{rel}$ as a function of time for 3-D logarithmic potentials with various $b_y$, where the axis of symmetry is perpendicular to the direction of motion.}
    \label{fig:8}
 \end{figure}

In the satellite described by a logarithmic potential, where the axis of symmetry is parallel to direction of motion, Figure \ref{fig:8}a, we find that the lowest value of the coefficient appears in the oblate spheroid $b_z=0.8$. For higher values of $b_z$ the coefficient is larger, however we notice, that while the prolate spheroid ($b_z=1.08$) has a higher $C_{rel}$, initially, it is then overtaken by the system with $b_z=0.9$. The spherically symmetric system with $b_z=1$ has the highest value of $C_{rel}$. Regarding the logarithmic potential satellites, for which the direction of motion is normal to the axis of symmetry of the system, Figure \ref{fig:8}b, the highest value of the $C_{rel}$ is in the prolate spheroid system with $b_y=1.08$, and is lower for smaller $b_y$. This noticeably different behaviour between the runs shown in Figures \ref{fig:8} a and b, is related to the direction of the velocity, where in the former it is parallel to the axis of symmetry of the potential whereas in the latter it is normal. Values of $b_z>1$ essentially correspond to values of $b_y<1$, so in principle, we see here a symmetry flipping. This affects the shape of the wake and consequently the value of $C_{rel}$. Finally, the results of the spherically symmetric cases ($b_z=1$, $b_y=1$) are identical.

\section{Discussion}

\label{DISCUSSION}

A key application of dynamical friction is related to the tendency of more massive objects to move towards the centre of the potential well and the overall mass distribution  \cite{2016JCAP...07..022A}. We note, however, that several systems such as Fornax dwarf spheroidal galaxies are expected to have dynamical friction timescales much shorter compared to their ages \cite{1998ApJ...501L..33B,1998MNRAS.297..517H,2003MNRAS.340..175M}, with a dependence on the DM distribution and the presence of a core or a central bulge  \cite{2006MNRAS.370.1829S,2008MNRAS.383...93B}. 

In our calculation we have explored values in the non-linear regime, with $M_Q<1$, which is the case for typical galaxies within clusters, whose velocities are in the range of $10^{3}$km/s \cite{2014ApJ...792...45R}.
We can further estimate the dynamical friction timescale for these sources using the expression \cite{Lancaster}:
\begin{equation}
    \tau_{DF} \sim \frac{v_{rel}^3}{4 \pi \bar{\rho}G^2 M C_{rel}}\,.
\end{equation}
Substituting the value for a DM density $\bar{\rho} \sim 10^6$M$_{\odot}$kpc$^{-3}$ \cite{2019MNRAS.484.1401R}, a velocity of $10^{3}$km/s and a mass of $10^{11}$M$_{\odot}$ and $C_{rel}$ we obtain a timescale similar to the Hubble time:
\begin{equation}
    \tau_{DF} \sim t_{H}\,.
\end{equation}
These results are consistent with previous calculations \cite{Lancaster}. However there is a substantial differentiation due to the ellipticity and the direction of motion, for two galaxies whose masses are equal, approximately by a factor of $5$ on the $C_{rel}$ coefficient. 

The effect of dynamical friction has been discussed in the context of galaxy clusters and infall of systems in the centres of clusters. There have been particular applications to Fornax \cite{2009ApJ...699.1389C}, a dwarf spheroidal galaxy surrounded by globular clusters distributed widely with respect to the galaxy \cite{2000ApJ...531..727O}, which should have been more closely distributed. More interestingly, the irregular shape of the Large Magellanic Cloud, makes the highly symmetric spherically symmetric systems inappropriate for the description of the dynamical friction wakes and deviations by up to a factor of a few may appear due to the different shapes. Moreover, in recent modelling of the impact of the LMC onto the DM distribution in the vicinity of the Milky Way, it is found that a wake is likely to appear \cite{2019ApJ...884...51G,2021ApJ...919..109G,2023MNRAS.tmp..536K}.

We note however, that, especially in the lower $M_Q$ models of FDM with non-spherically symmetric satellites, as is the case with LMC, the formation of DM underdensitites and finer structure configurations in the halo will be expected. Moreover, an interesting finding is the reversal of the dynamical friction. This result has been reported also in simulations of the dynamical friction near the central soliton region of the FDM halo, through self-consistent solution of the Schr\"odinger-Poisson equation \cite{2021ApJ...916...27D}, where the entire motion is affected and the satellite oscillates. Similar behaviour has been found in perturbed dark matter halo which instead of generating a friction force, can provide buoyancy and actually push the satellite away from the critical radius of the core instead of attracting it \cite{2021ApJ...912...43B, 2022ApJ...926..215B}. Simulations of the dynamical friction due to FDM in the relativistic regime lead to an oscillatory behaviour of the force, as well. This force, while on average opposes the motion it can lead to short intervals of  acceleration  \cite{2021PhRvD.104j3014T}. 
Finally, an interesting comparison can be made with studies of the dynamical friction due to a gaseous medium \cite{1999ApJ...513..252O}. The dynamical friction exerted by the gaseous halo is more efficient than the classical estimation of Chandrasekhar for supersonic flows and less efficient for subsonic ones. In the FDM simulations studied here, we notice that the dynamical friction coefficient exceeds unity in the low quantum Mach number regime, and fluctuates between positive and negative values leading to both drag and acceleration.

While here we assess the impact of the satellite onto the FDM halo, an interesting question would be to relax our initial choice of satellite potential and let it evolve in time. While this problem is more complex in terms of the choice of the initial condition, its numerical implementation requires minimal modifications on the existing numerical framework. Furthermore, an interesting open question would be the exploration of other geometries that have used in modelling of galactic and globular cluster potentials, in order to quantify particular systems and assess the validity of the proposed models.

\section{Conclusions}
\label{CONCLUSIONS}

In previous works related to the interaction of a mass with FDM, the emphasis is placed on spherically symmetric potentials. Most large-scale structures however, especially Galaxies, are unlikely to be spherically symmetric and are typically described by axisymmetric or even triaxial potentials. In this paper, we have performed numerical simulations with particular focus to logarithmic axisymmetric potentials. We notice significant qualitative differences between spherically symmetric satellites and axisymmetric ones, even if the other parameters are identical. In particular we notice that the shape of the wake and the overall deformation the satellite creates to the FDM halo is strongly dependent on its ellipticity. Such features may be identifiable in future high resolution maps of DM through gravitational lensing and provide a distinction between the various proposed DM models. 

Satellites with the same ellipticities have different dynamical friction coefficients, depending on their direction of motion: whether it is along the symmetric axis or normal to it. Consequently, the dynamical friction exerted depends on these characteristics, and it is likely two similar satellites to experience gravitational frictions differing by a factor of 5 because of this. The combination of satellite masses, dark matter average densities, and relative velocities, places typical members of galaxy cluster in the low $M_Q$ regime, which is, however, non linear and leads to differences in the wake formation compared to the analytical theory. We have also found, in accordance with previous works \cite{Lancaster}, that in the limit of low velocities, the appearance of DM overdensities and underdenisities may lead, temporarily, to the gravitational acceleration rather than drag.

\acknowledgments
We would like to thank P. Mocz, L. Lancaster, L. Hui, A. Gruzinov and G. Gelmini for discussions during the preparation of this manuscript. We also want to thank an anonymous referee for her/his insightful comments that improved this manuscript. 

KNG acknowledges funding from grant FK 81641 "Theoretical and Computational Astrophysics", ELKE.

The authors acknowledge COST ACTION Cosmic WISPers CA21106 for inspiration on this topic.

\bibliography{mybib}

\end{document}